\documentclass{cpbtex}

\usepackage{amsmath}
\usepackage{amssymb}
\usepackage{graphicx}
\usepackage{hyperref}
\usepackage{url}
\usepackage{color,xcolor}
\usepackage{epstopdf}
\usepackage{float}
\usepackage{ulem}
\usepackage{multirow}
\usepackage{amsmath,bm}
\makeatletter
\newcommand{\figcaption}{\def\@captype{figure}\caption}
\newcommand{\tabcaption}{\def\@captype{table}\caption}
\makeatother

\begin{document}

\title{Ferroelectricity induced by the absorption of water molecules
         on double helix SnIP}


\author{Dan Liu$^{1}$\thanks{Corresponding author.
        E-mail:~liudan2@seu.edu.cn}, Ran Wei$^{1}$, Lin Han$^{1}$, Chen Zhu$^{1}$ and Shuai Dong$^{1}$\\
$^{1}${School of Physics,
       Southeast University,
       Nanjing 211189, China}}  


\date{\today}
\maketitle

\begin{abstract}
We study the ferroelectricity in a one-dimensional system composed of
a double helix SnIP with absorbing water molecules. Our
$\it{ab~initio}$ calculations reveal two factors that are
critical to the electrical polarization. The first one is the
orientation of polarized water molecules staying in the R2 region
of SnIP. The second one is the displacement of I atom which roots from subtle interaction with
absorbed water molecules. A reasonable scenario of polarization
flipping is proposed in this study. In the scenario, the water
molecule is rolling-up with keeping the magnitude of its electrical
dipole and changing its direction, meanwhile, the displacement of I
atoms is also reversed. Highly tunable polarization can be
achieved by applying strain, with $26.5\%$ of polarization
enhancement by applying tensile strain, with only $4\%$ degradation
is observed with $4\%$ compressive strain. Finally, the direct band
gap is also found to be correlated with strain.
\end{abstract}

\textbf{Keywords:} ferroelectricity, 1D double helix, electrical polarization, DFT

\textbf{PACS:} 77.80.-e, 77.84.-s, 73.22.-f

\section{Introduction}

Since the first experimental discovery of ferroelectricity
in Rochelle salt~\cite{Triscone07}, in the past hundred years,
for the perspective of application in sensors, actuators and
memories, a wide range of ferroelectric (FE)
materials have been intensively studied, such as the
well known three dimensional (3D) BaTiO$_{3}$~\cite{Eom2004}.
Interest in new, low-dimensional ferroelectrics has also been
growing rapidly due to their potential applications in electronics,
such as recently emerged ferroelectric two-dimensional (2D) layered
structures including group-VI monochalcogenides, In$_{2}$Se$_{3}$,
CuInP$_{2}$S$_{6}$,
etc~\cite{{Yang2016},{Xue2017},{Wang2017},{Duan2020},{Zhu2017},{Liu2016},{Xiao2018},{Wu2017},{Dong2021},{Guan2021},{Ziwen2021},{Ding2020},{Lin2019},{You2019}}.
Besides these intrinsic 2D FE materials, ferroelectricity can be
also induced in non-FE 2D system through doping,
manipulating defects and also fabricating
heterojunctions. For example, the hydroxyl-decorated
graphene systems exhibit a robust
ferroelectricity~\cite{{Wu2013},{Kan2013},{Lin20191}}. The
increasing demand in high density in electronic device requests the
further miniaturization of FE materials. One method is to
cut a one-dimensional (1D) nanoribbon from the traditional 3D or
2D FE materials. However, there are two obvious drawbacks of
these 1D nanoribbons. One is the size of such nanoribbon is quite
large. For example, graphene nanoribbons fabricated
by etching graphene using high-resolution electron beam lithography
range from 10~nm to 100~nm~\cite{Hernandez12}. This size is much
larger than typical 1D structures, such as the (10, 10) carbon
nanotube (CNT) of which the diameter is only about
1.3~nm~\cite{Iijima93}. So these nanoribbons can not be
categorized as real 1D ferroelectricity, and thus the
reduction effect in electronic device size using these FE
nanoribbons is limited. The second concern is about the
stability of these nanoribbons, while cutting from their 3D or 2D
parent compounds. The atoms staying on the edge are
electrical unsaturated, which will result in a structural
distortion and accordingly the FE behavior of these nanoribbons.
In general, the optimum method to reduce the size and increase
the density of electronic devices should be seeking real 1D
ferroelectricity in atomic scale. Despite abundant 1D structures
that have been explored since the successful synthesis of
CNT~\cite{Iijima91}, intrinsic 1D FE materials are rarely
found~\cite{{Ren2021},{Yakob2019},{Du2021}}. In analogy with
decorating graphene using hydroxyl to introduce electrical
polarization, a similar effect may be expecting in existed 1D
structures through doping of polarized molecules.

\begin{center}
\includegraphics[width=0.8\columnwidth]{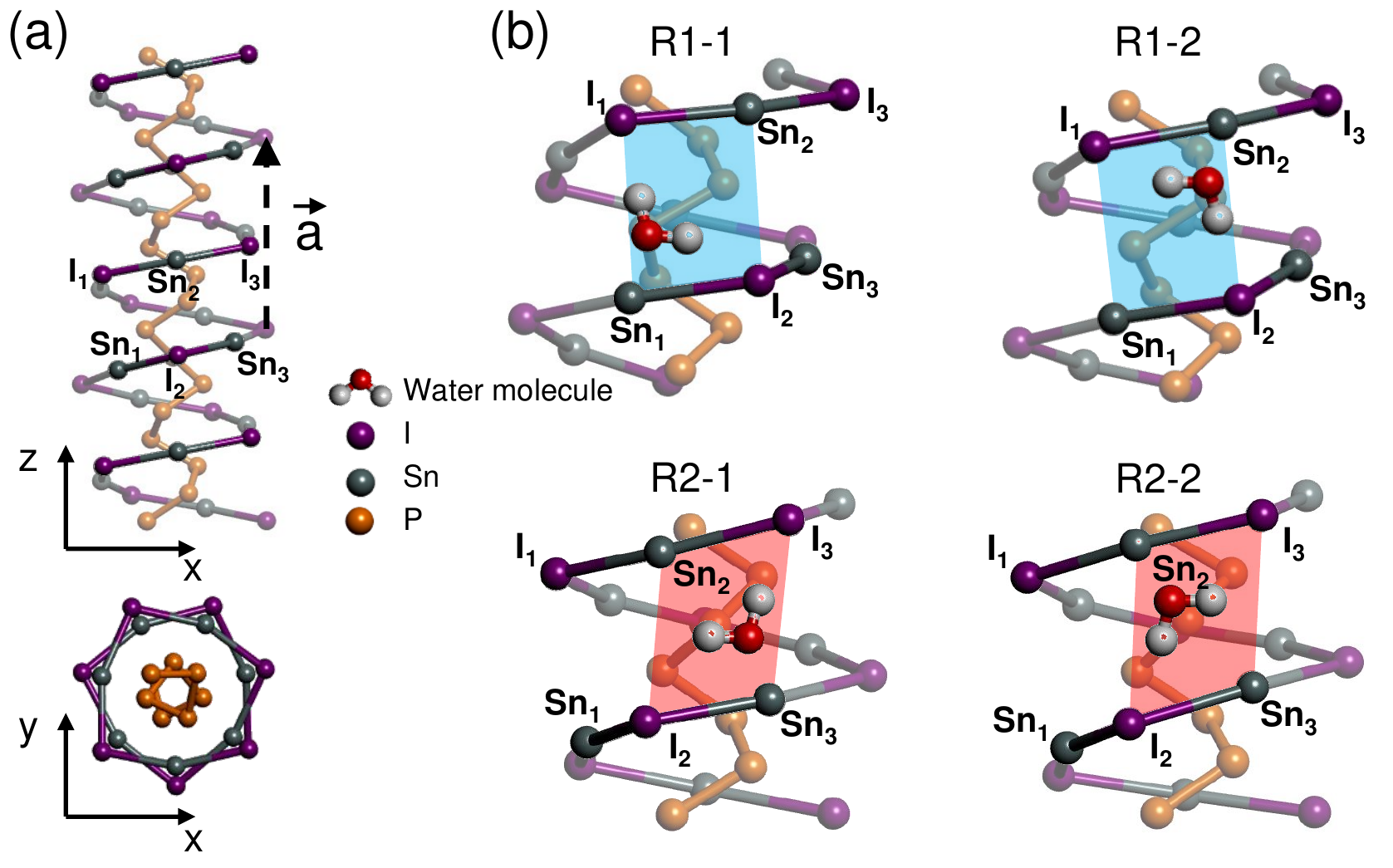}\\[5pt]  
\parbox[c]{15.0cm}{\footnotesize{\bf Fig.~1.}
Atomic structure of (a) 1D double-helix SnIP, (b) four
configurations of one water molecule attaching on region R1 of
SnIP highlighted by transparent blue area, on region R2
highlighted by red transparent red area. Black dashed arrow
represents the vector of unit cell. %
\label{fig1}}
\end{center}

Water molecule is a very common polarized molecule around us.
However, little attention has been focused on the ferroelectricity
of water to date, because it is not easy to make water molecules
form a well-organized arrangement. To solve the problem, we need to
find an appropriate way to enforce water molecules to form an
expected arrangement. Previous study reported a unique structure of
quasi-1D water molecules confined in a 3D supramolecular
architecture. When the water molecules are frozen into special
forms, this quasi-1D water wire becomes
ferroelectric~\cite{Zheng2011}. In this study, we investigate the
electric polarization in 1D double-helix
SnIP~\cite{{Nilges2016},{Hoff2021}} with absorption of water
molecules. Our density functional theory (DFT) calculations reveal
that there are two regions R1 and R2 in SnIP to anchor water
molecules. And for each region, there are also two energetic
degenerate configurations holding promise of the reversal of
water molecule. We find that there are two factors contributing to
the emergence of ferroelectricity in the system. One is the
orientation of the polarized water molecules, the other is the
displacement of ions in SnIP originating from the interaction
between water molecules and SnIP. While flipping the orientation of
water molecules from one degenerate configuration to the other, the
displacement of ions of SnIP will also be reversed, which
leads to the reversal of an entire electrical dipole. The
magnitude of polarization can be tuned mainly through applying
tensile strain along the axis of SnIP, while
almost remaining constant under compressive strain. The band
gap of the system, remaining to be direct, can be tuned by the
axial strain as well.

\section{Theoretical method}

Our calculations of the stability, equilibrium structure,
polarization and energy changes during structural transformations
have been performed using the density functional theory (DFT) as
implemented in SIESTA code~\cite{Artacho2008}. The 1D systems have
been represented using periodic boundary conditions and separated
by 20{\AA}. We have used the nonlocal Perdew-Burke-Ernzerhof
(PBE)~\cite{Perdew1996} and Local-Density-Approximation
(LDA)~\cite{Ceperley1980} exchange-correlation functional,
norm-conserving Troullie-Martins
pseudopotentials~\cite{Troullier1991}, and a local numerical
double-$\zeta$ basis including polarization orbitals. The Brillouin
zone of periodic structures has been sampled by a fine grid of
1$\times$1$\times$12 k-points for 1D structures~\cite{Pack1976}. We
find the basis, k-point grid, and mesh cutoff energy of 180 Ry
used in the Fourier representation of the self-consistent charge
density to be fully converged, providing us with a precision in
total energy of 2 meV/atom. Geometries have been optimized using
the conjugate gradient (CG) method~\cite{CGmethod}, until none of
the residual Hellmann-Feynman forces exceeded $10^{-2}$~eV/{\AA}.
For the phonon spectrum calculation, we use a much smaller force
tolerance of $10^{-3}$~eV/{\AA} to get the optimized structure.
The polarization is calculated using Berry phase
method~\cite{{Resta1993},{Resta2007}}.

\section{Results and Discussion}

\begin{center}
\includegraphics[width=0.8\columnwidth]{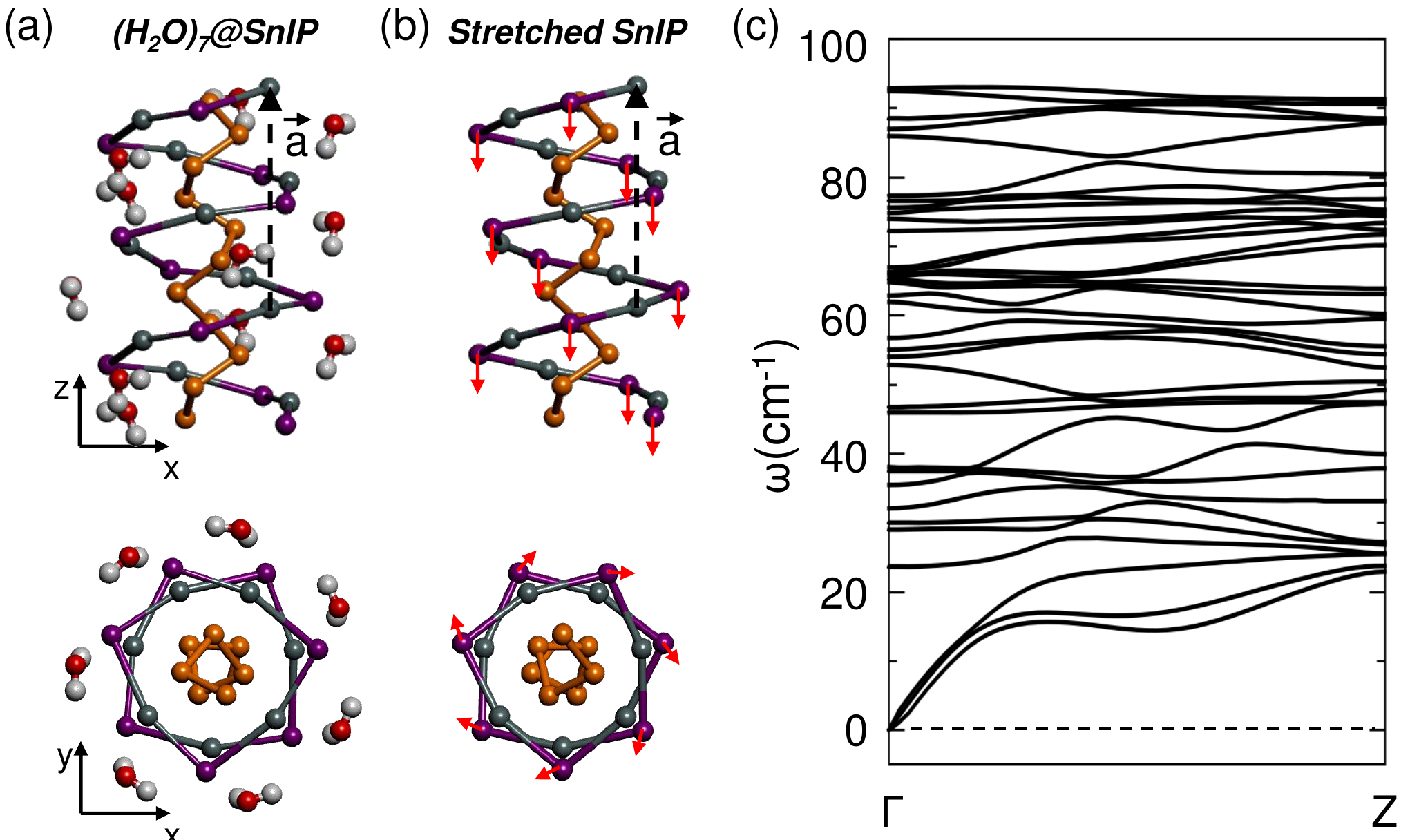}\\[5pt]  
\parbox[c]{15.0cm}{\footnotesize{\bf Fig.~2.}
Atomic structure of (a) (H$_{2}$O)$_{7}$@SnIP with 7 H$_{2}$O
attaching on SnIP, (b) stretched SnIP with the same unit cell size
of (H$_{2}$O)$_{7}$@SnIP in (a). (c) Phonon spectrum of
(H$_{2}$O)$_{7}$@SnIP. The red arrow with solid line represent the
shift of the atoms transforming the stretched SnIP in (b) to SnIP in
(H$_{2}$O)$_{7}$@SnIP of (a).%
\label{fig2}}
\end{center}

\subsection*{Atomic structure of 1D double-helix SnIP and
             water@SnIP}

The bunch of SnIP has been synthesized successfully in
experiment~\cite{Nilges2016}. The single rod of double-helix SnIP
is composed of the inner Phosphorus (P) helix and the outer
Tin(Sn)-Iodine(I) helix, which contains seven P, Sn and I atoms in
one primitive unit cell. The double helix of SnIP can be either
right-handed or left-handed with the same energetic stability. Here, we
focus on the right-handed SnIP to study the electrical
polarization, since the left-handed SnIP will possess the
same FE properties as the right-handed one. Our DFT calculation
indicates the lattice constant of $\vec{a}$ as shown in Fig.~1(a)
is 8.12~{\AA} which is slightly larger than the experimentally
observed value of bunched SnIP by 0.19~{\AA}~\cite{Nilges2016}.
Considering the interaction between each rod in the bunch will
reduce the length along $\vec{a}$, our calculation result is quite
reliable. We analyze the Mulliken population of each atom of SnIP,
and find that Sn carries 0.45 positive charges while P and I get
0.22 and 0.23 more electrons respectively. This result clearly
shows SnIP is an ionic compound, and it could be a potential
candidate of FE materials. However the space group of $P2$ of SnIP
shows a $180^\circ$ rotation symmetry in the direction along the
axis perpendicular to the helix, which prohibits the polarization
in this pristine system. Further modification of the structure is
needed to induce ferroelectricity.

\begin{center}
\includegraphics[width=0.8\columnwidth]{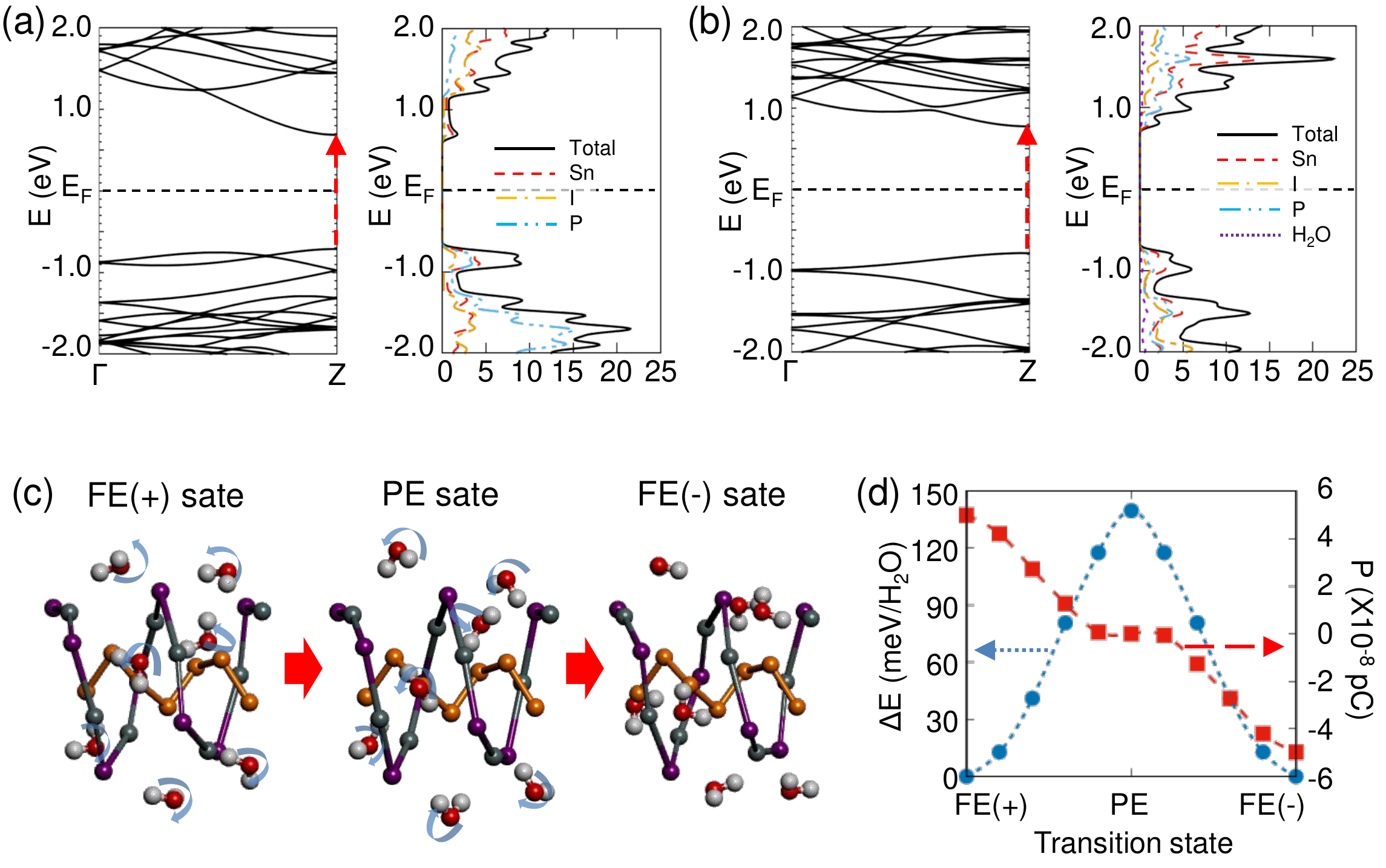}\\[5pt]  
\parbox[c]{15.0cm}{\footnotesize{\bf Fig.~3.}
Electronic band structure ($E_{k}$) and density of states (DOS)
including its projection on individual atoms (left panels) for (a)
optimum 1D double-helix SnIP, (b) (H$_{2}$O)$_{7}$@SnIP. (c)
Illustration of flipping ferroelectricity of (H$_{2}$O)$_{7}$@SnIP,
FE(+) sate and FE(-) state are energetic eigenvalue states with
opposite polarization direction. PE state is the intermediate
paraelectric state. (d) The change of energy and polarization in
the process of transformation from FE(+) to FE(-) state.%
\label{fig3}}
\end{center}

An in-depth analysis of the double-helix SnIP reveals that for the
outer helix SnI, there are two types of regions namely $R1$
and $R2$, which are highlighted by blue and red shadows respectively
in Fig.~1(b). In one primitive unit cell of SnIP, there are seven
alternatively arranged $R1$ and $R2$ regions. For each region,
it is surrounded by two Sn atoms and two I atoms. $R1$ is enclosed
by I$_{1}$-Sn$_{1}$-I$_{2}$-Sn$_{2}$, and $R2$ is enclosed by
Sn$_{2}$-I$_{2}$-Sn$_{3}$-I$_{3}$. In $R1$, the distance $d$
between two I atoms $d$(I$_{1}$-I$_{2}$)=4.92~{\AA}, and $d$
between two Sn atoms $d$(Sn$_{1}$-Sn$_{2}$)=5.41~{\AA}, while in
$R2$, $d$(I$_{3}$-I$_{2}$)=5.81~{\AA} and
$d$(Sn$_{3}$-Sn$_{2}$)=4.42~{\AA}. These two different
regions offer two different potential spaces. Then we investigate
the absorption of water molecules on the surface of the outer helix
SnI. We first put one water molecule in $R1$ and $R2$ to see which
region the water molecule prefers to stay. When water
molecules are absorbed in $R1$ or $R2$, the two hydrogen (H) atoms
of water molecule are attracted by two I atoms, and the oxygen (O)
atom tilts to one of the two Sn atoms. In other words, $R1$ and
$R2$ provide a double-well potential for water molecules. As a
result, there are two configurations for one water molecule
attached in $R1$ or $R2$, labeled by $R1-1$, $R1-2$, $R2-1$ and
$R2-2$ shown in Fig.~1(b). Our DFT-PBE calculations indicate that
the cohesive energy of ${\Delta}E_{PBE}(R1-1)=275$~meV,
${\Delta}E_{PBE}(R1-2)=277$~meV, ${\Delta}E_{PBE}(R2-1)=298$~meV
and ${\Delta}E_{PBE}(R2-2)=299$~meV. Taking the DFT calculation
error of energy into consideration, we can conclude that the
configuration of $R1-1$ is as energetically stable as $R1-2$ and
$R2-1$ is as energetically stable as $R2-2$. Considering SIESTA
uses pseudo-atomic-orbital basis sets, the cohesive energy will be
overestimated due to the basis set superposition error (BSSE). We
have done correction calculations for BSSE (BSSEc), and find that
${\Delta}E_{PBE-BSSEc}(R2)=150$~meV and
${\Delta}E_{PBE-BSSEc}(R1)=135$~meV. We also use DFT-LDA to
calculate the cohesive energy with BSSE correction taken in
account, and find that ${\Delta}E_{LDA-BSSEc}(R2)=426$~meV and
${\Delta}E_{LDA-BSSEc}(R1)=398$~meV. Both the results of DFT-PBE
and DFT-LDA show that the configuration of water molecule staying
in $R2$ is more stable than in $R1$. From previous
works~\cite{{Kyuho07},{DT256}}, we should claim here that the real
cohesive energy of ${\Delta}E(R2)$ should be between 150~meV and
426~meV. The mulliken charge analysis of configurations in $R2$
shows a charge transfer from water molecule to SnIP with $0.03$~e.

Then we investigate two molecules absorbed in two neighboring $R2$
regions of SnIP separately. We find that the two water molecules
aligning in the same direction are more stable than aligning in the
opposite direction by 34~meV. Next, we focus on the water
molecules aligning in the same direction while staying on SnIP. As
mentioned above, in one primitive unit cell of SnIP,
there are seven $R2$ regions available for water molecules to stay.
We assume the simplest case that each $R2$ region only absorbs one
water molecule. The formed chemical formula is written as
(H$_{2}$O)$_{7}$@SnIP as the atomic structure shown in Fig.~2(a).
The cohesive energy of absorbing seven water molecules in $R2$
regions in one unit cell is ${\Delta}E(R2)=295$~meV. The cohesive
energy ${\Delta}E(R2)$ is stronger than a typical van der Waals
interaction, which not only locks the molecule in a specific
orientation, but also results in the distortion in atomic
structure of SnIP. After absorbing seven water molecules in one
unit cell, the system is elongated by $4.4\%$. The dynamic
stability of (H$_{2}$O)$_{7}$@SnIP is confirmed by the phonon
spectrum shown in Fig.~2(c).

\begin{center}
\includegraphics[width=0.8\columnwidth]{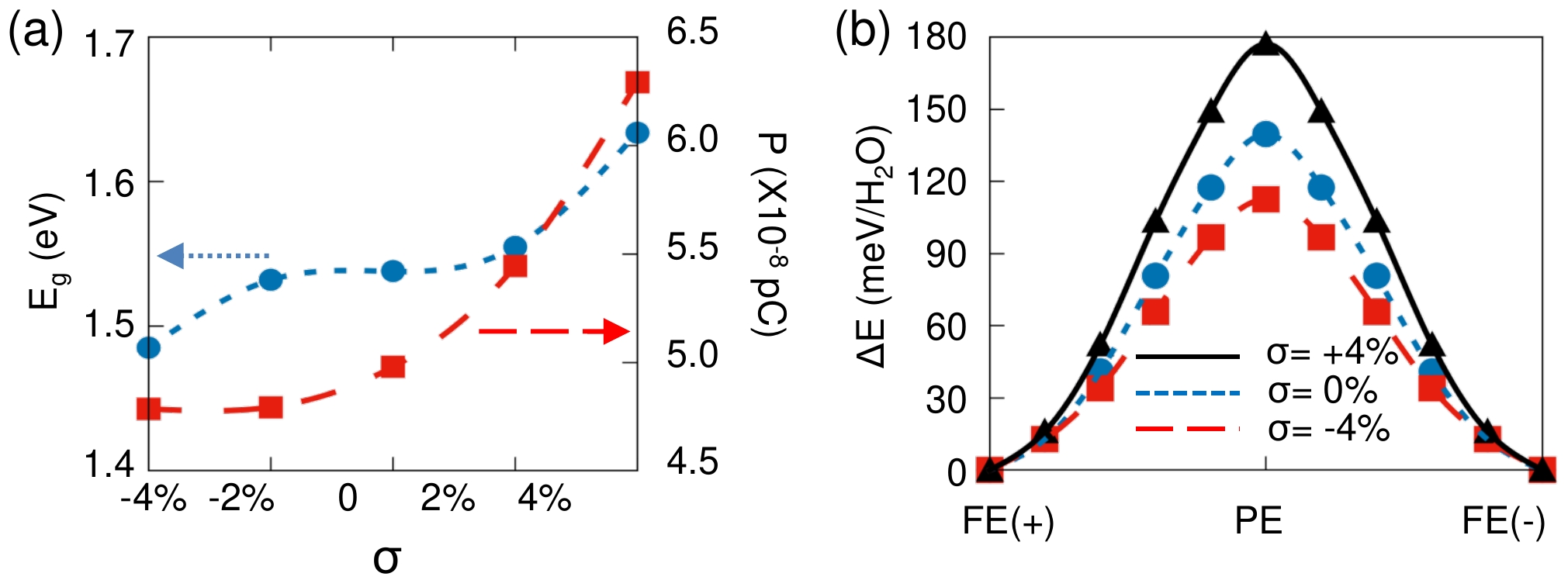}\\[5pt]  
\parbox[c]{15.0cm}{\footnotesize{\bf Fig.~4.}
(a) Band gap value $E_{g}$ and the polarization as function of the
axial strain. (b) The change of energy in the process of
transformation from FE(+) to FE(-) state with axial strain of $-4\%$,
$0\%$ and $+4\%$.%
\label{fig4}}
\end{center}

\subsection*{Band structure and ferroelectric behavior of (H$_{2}$O)$_{7}$@SnIP}

The isolated SnIP is a semiconductor with a direct band gap of
$1.32$~eV at $Z$-point shown in Fig.~3(a). We should claim here
that the bandgap is underestimated by DFT-PBE calculation, but
still provides useful insights into the trend in the
electronic structures. For (H$_{2}$O)$_{7}$@SnIP, a direct
band gap of $1.48$~eV at $Z$-point is shown in Fig.~3(b). The
dispersion relation at the top of valence bands and bottom of
conduction bands is similar for both systems, which may come
from SnIP but not water molecule. This is confirmed by the partial
density of state in Fig.~3(a) and (b), where we can see that the
contribution from water is almost zero in the energy range from
$-2$~eV to $2$~eV. So while the SnIP is stretched after absorbing
water molecules, the orbital hopping of SnIP is reduced, which
results in an increase of bandgap by $0.16$~eV.

To study the ferroelectricity, we first analyze the detailed shift
of each atoms of the system. After combining water molecules and
SnIP, besides the oriented arrangement of water molecules and the
stretch of SnIP resulted by the interaction, there is also a
local displacement of atoms. We compare the atomic structure of
SnIP in (H$_{2}$O)$_{7}$@SnIP with the isolated SnIP stretched by
$4.4\%$. We find an obvious displacement of I atoms by $0.3$~{\AA}
along the axis direction, while the displacement of Sn and P atoms
can be ignored. The illustration of the shift of I atoms is
represented by the red arrow in Fig.~2(a). Besides the displacement
of I atoms
along the axis direction, there is also a slight shift of I atoms
in the plane perpendicular with the axis. These distortions of
atomic structure make the space group $P2$ of pristine SnIP change
to space group $P1$ of SnIP in (H$_{2}$O)$_{7}$@SnIP.


The ferroelectricity of the system can be divided into two main
parts. The first one is the arrangement of the dipole of water
molecules. Since the H$_{2}$O is an ionic molecule, the
polarization of water molecule $P$(H$_{2}$O) can be defined as the
charge of O ion times the distance between the center of positive
and negative charge. Since the center of positive and negative
charge is overlapped in the pristine SnIP, there is no net
polarization. After absorbing the water molecules, we see the
displacement of I ion carrying $0.22$ negative charges, then
$P(SnIP)$ is defined as the charge of I ion times the shift of I
ion. $P_{Z}$ is the decomposition of $P$ in the Z direction. In the
DFT calculation, we usually use berry phase method installed in the
software to calculate the polarization directly. From Fig.~2(a), we
see the component of water dipole along axis $P_{Z}$(H$_{2}$O) is
in the negative $Z$-direction, and the shift of I atoms in the
negative $Z$-direction results in an electric dipole $P_{Z}$(SnIP)
in the positive $Z$-direction. If we do not consider the charge
redistribution between atoms caused by absorbing water molecule on
SnIP, the total net electric polarization $P_{Z}$ should be the
difference between $P_{Z}$(SnIP) and $P_{Z}$(H$_{2}$O). We first
separate the 1D water wire and the SnIP, and then calculate
$P_{Z}$. We find that $P_{Z}$(H$_{2}$O)=$3.61\times 10^{-8}$~pC and
$P_{Z}$(SnIP)=$6.53\times 10^{-8}$~pC respectively, so
$P_{Z}=2.92\times 10^{-8}$~pC. However, the charge redistribution
can not be neglected, and it will modify $P_{Z}$(H$_{2}$O) and
$P_{Z}$(SnIP). We study the charge $Q$ of atoms before and after
absorbing water molecules on SnIP. $Q$(O) decrease from -0.78 to
-0.73, $Q$(H) of half of H atoms decrease from 0.39 to 0.36, while
the charge of other half keeps the same, the decrease of charges of
H and O results in the decrease of electric dipole of the water
molecule. Instead, $Q$(Sn) increases from 0.45 to 0.47 and $Q$(I)
increases from -0.23 to -0.27, which makes an increase of
dipole moment of SnIP. As a result, the increase of $P_{Z}$(SnIP) and
the decrease of $P_{Z}$(H$_{2}$O) lead to a larger $P_{Z}$ than
$2.92\times 10^{-8}$~pC. In order to get a reliable result, we use
the standard Berry phase method and get $P_{Z}=4.98\times
10^{-8}$~pC, which confirms the effect of charge redistribution on
the polarization. This electrical polarization of
(H$_{2}$O)$_{7}$@SnIP is comparable with the value in BaTiO$_{3}$
of $3.2\times 10^{-8}$~pC~\cite{Ren2004}. Besides the shift along
the axial direction, I atoms also shift in the plane perpendicular
to the axis, however these shift are so small that the resulted
polarization is negligible.

\begin{center}
\includegraphics[width=0.8\columnwidth]{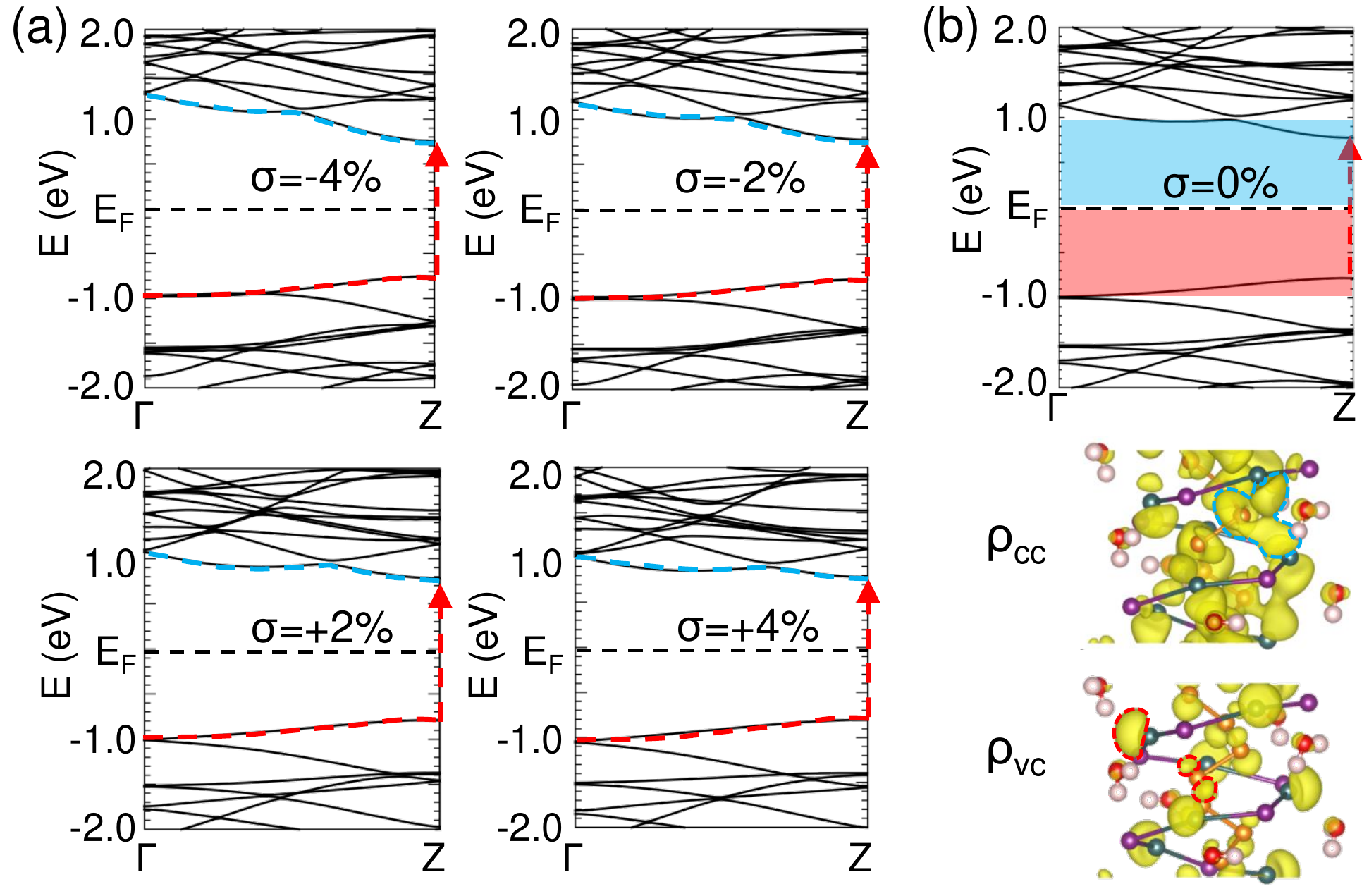}\\[5pt]  
\parbox[c]{15.0cm}{\footnotesize{\bf Fig.~5.}
(a) Band structure $E_{k}$ of (H$_{2}$O)$_{7}$@SnIP under axial
strain. The top valence band is marked by red dashed line and the
bottom conduction band is marked by blue dashed line. (b) Charge
density associated with valence state $\rho_{vc}$ and conduction
state $\rho_{cc}$ of optimum (H$_{2}$O)$_{7}$@SnIP is present in the
up and bottom panel respectively. The energy range
associated with $\rho_{vc}$ is indicated by the red transparent
region in the band structure of optimum (H$_{2}$O)$_{7}$@SnIP extends
from $E_{F}-0.94<E<E_{F}$, and the energy range associated with
$\rho_{cc}$ is indicated by the blue transparent region in the band
structure of optimum (H$_{2}$O)$_{7}$@SnIP extends from
$E_{F}<E<E_{F}+0.94$. The isosurface values of $\rho_{vc}$ and
$\rho_{cc}$ are $1{\times}10^{-3}$e/{\AA}$^3$ and
$0.5{\times}10^{-3}$e/{\AA}$^3$.%
\label{fig5}}
\end{center}

While applying an external electric field in the opposite direction
with the polarization of a FE material, the polarization can be
turned over by $180^\circ$. In the process of flipping the
polarization, the system starts from the initial FE state FE(+),
goes through a PE phase in which there is no net polarization, and
arrives at the final FE state FE(-). FE(-) has an opposite
polarization but the same energetic stability as FE(+). In the
traditional 3D or 2D FE materials like BaTiO$_3$, the PE phase
is typically around the midpoint of a linear combination of the FE(+)
and FE(-). However, in the PE phase acquired through this way in
our system, the bond angle of H-O-H in the water molecule is around
$180^\circ$, far from the optimum bond angle. As
expected, the energy of PE phase is much higher than FE phase by
4.65~eV/H$_2$O, which means the PE phase obtained through linear
combination is not acceptable.

We propose here another reasonable scenario of flipping the
electrical polarization as shown in Fig.~3(c). In this scenario,
the water molecule is rolling-over while keeping the dipole
constant, but changing the direction. In the PE phase, the dipole
of water molecule lies in the radial direction of the double helix.
In other words, there is no decomposition of dipole of water in the
axis direction. In the process of rolling-over of water molecules,
due to the interaction between water molecule and SnIP, the atomic
structure of SnIP is also changing. In the PE phase, the shift of I
atoms and also the associated polarization disappears. As a
result of vanishing of the polarization of water molecule and SnIP,
there is no net polarization of the whole system in the axis
direction. As mentioned above, we know the adjacent water molecules
staying in R2 regions prefer to align in the same direction. In the
process of transformation from FE(+) to FE(-), once one water
molecule starts to turn over, the other neighboring water molecules
will follow to turn over. In other words, the seven water molecules
flip collectively in the transformation process. The changes of
energy and net polarization along the axis in the process of
flipping the polarization are shown in Fig.~3(d). The energy
barrier is about 140~meV/H$_2$O which is about thirty-three times
smaller than that of the linear combination scenario.

\subsection*{Strain effect on Ferroelectricity and band structure}


We also study the response of ferroelectricity and band structure
to the axial strain to explore the perspective of the materials in
application. The response of $P_{Z}$ to the tensile
strain is more obvious than the compressive strain. From red
dashed line in Fig.~4(a), we see $P_{Z}$ increases
significantly by $26.3\%$ with $4\%$ stretch strain, but decreases
very slightly by $4.4\%$ with $4\%$ compress strain. The band gap
$E_{g}$ of the system is also decreased to $1.49$~eV with the $4\%$
axial compression and increased to $1.63$~eV with $4\%$ axial
stretching. The effect of strain on the band gap is shown with blue
dotted line in Fig.~3(c). The deformation energy ${\Delta}E$ with
strain of $4\%$ is about 70~meV, which means this double-helix can
be deformed easily by applying an axial stress. The energy
barrier ${\Delta}E$ in the process of transformation from FE(+) to
FE(-) is also affected by the strain. As shown in Fig.~4(b), from
$4\%$ compressive strain to $4\%$ tensile strain, ${\Delta}E$ is
increased from 112~meV/H$_2$O to 177~meV/H$_2$O.

The change of the band gap with applying strain may come from the
change of band width. In order to explore the physics behind, we
calculate the band structures of (H$_{2}$O)$_{7}$@SnIP with
different stress conditions as shown in Fig.~5(a). Comparing
with the band structure of the system without strain, we find the
top valence band highlighted by the red dashed line almost
remains unchanged. The width of bottom conduction band highlighted
by blue dashed line changes but still keeps the dispersion
relation. In Fig.~5(b), the energy range between $E_{F}$ and 0.2~eV
below the top of valence band indicated by the red region, and the
0.2~eV above the bottom of conduction band indicated by the blue
region, are used to identify the valence and conduction frontier
states. We characterize the charge density associated with valence
frontier state $\rho_{vc}$ and conduction frontier state
$\rho_{cc}$ in the lower panel of Fig.~5(b). The valence frontier
states encircled by red dashed line is mainly occupied by the lone
pair orbital of Sn and $p_{z}$ orbital of P. The conduction
frontier states encircled by blue dashed line contain the overlap
of orbital of one P atom and two nearest-neighboring Sn atoms.
While applying strain on the system, the distances between atoms
are changed. Since the valence frontier state has little relation
with overlap of orbital of atoms, there should be little change of
the top valence band. However, the change of distance affects the
overlap of orbital between P and Sn atoms, so we see the change of
conduction band. While stretching the system, the distance between
two nearest-neighbor Sn atoms is enlarged from $4.54$~{\AA} to
$4.61$~{\AA}, and the distance between P atom and its
nearest-neighbor Sn atom is enlarged from $3.46$~{\AA} to
$3.56$~{\AA}. As a result, the overlap between orbital and the
resulted width of conduction band is decreased, and vice versa for
the compressive strain.

\subsection*{Discussion}


When we study the ferroelectric behavior, band structures and their
response of axial strain in 1D (H$_{2}$O)$_{7}$@SnIP, one basic
question may arise. Is SnIP hydrophilic or hydrophobic and
could SnIP attract water molecule from a water cluster? The answer
is determined by the competition between the absorption
energy of water molecule on SnIP and the interaction between water
molecules in water cluster. There are many configurations of water
clusters, such as the well known water dimmer. Using PBE-BSSEc, we
calculate the interaction between water molecules $E_{PBE-BSSEc}$
in the dimmer configurations. It ranges from $46$~meV to $246$~meV.
In our system, the interaction energy between water and SnIP is
about $150$~meV, which holds a great promise for water
molecule to stay.

Usually, for a semiconductor, while using strain to improve one
physical property, other properties will be degraded. In this
study, we find that the modification of polarization by axial
strain is very prominent, while the effect of strain on band
structure including the dispersion relation and band gap is
negligible. So with strain, the polarization is changed while other
physical properties are intact. Besides the strain along
axis direction, the other common strain occurs frequently in 1D
system is bending. In theoretical models of DFT calculation, it
will contain too many atoms to study the bending effect on physical
properties. However, we can analysis the bending effect
qualitatively through the studying effect of axial
tensile/compressive strain. The diameter of 1D
(H$_{2}$O)$_{7}$@SnIP is about $8.2$~{\AA}, which is the
same as (6,6) carbon nanotube. When applying the bending strain,
the outer part of this 1D double-helix is stretched and the inner
part is compressed. If we consider the bending 1D tube as a part of
a circle, then the extent of stretch of the outer part is the same
with the extent of compression of the inner part. From Fig.~3(c),
the response of polarization to tensile strain is much more
significant than the response to compressive strain. As a result,
we can conclude that the polarization will also be increased
by bending the 1D (H$_{2}$O)$_{7}$@SnIP.

The special atomic structure of SnIP makes it to be a promising
candidate to explore ferroelectricity in 1D. Here we use water
molecule to introduce electrical polarization in SnIP. The double
potential well in $R2$ of SnIP enforces the water molecule to
arrange in a special orientation. We can use other polarized
molecules to replace water molecules, and it will have similar
results. In addition, we can use transition metal atoms to dope SnIP.
With appropriate choice of transition metal element, the atoms will
stay in one of the double-potential wells of SnIP to generate the
ferroelectricity. Besides, the doping of transition metal element
will also bring magnetism to the 1D system.


\section{Conclusion}

In conclusion, we have identified ferroelectricity induced in 1D
direct-gap semiconductor (H$_{2}$O)$_{7}$@SnIP through absorbing
water molecule in SnIP. Our $\it{ab~initio}$ DFT calculations
indicate that there are two double-potential well regions $R1$ and
$R2$ of SnIP for water molecule to stay, and water molecule prefer
to attach in $R2$. The water molecules are arranged in a special
orientation due to the double-potential well of $R2$. On the other
hand, the interaction between water molecule and SnIP will lead to
the distortion of the atomic structure of SnIP and also a charge
redistribution between atoms. Comparing with the pristine SnIP with
space group of $P2$, (H$_{2}$O)$_{7}$@SnIP with space group of $P1$
loses the $180^\circ$ rotation symmetry. For polarization
flipping mechanism, we propose a scenario in which the water
molecules rotate while keeping the magnitude of the dipole but
changing the direction. The rotation of water molecule also leads to
the deformation of atomic structure of SnIP. When the dipole of
water molecule lies in the radial direction, SnIP is transformed
back to the phase with space group of P2. As a result, the net
polarization along the axis direction disappears. The effect of
external axial strain on polarization is quite remarkable,
with 4$\%$ stretching, the polarization is enhanced by about
26$\%$. However the response of $E_{g}$ of the band structure to
the axial strain is not significant, since only the bottom of
conduction band, instead of top of the valence band,
correlates with orbital overlap of atoms.

\addcontentsline{toc}{chapter}{Acknowledgment}
\section*{Acknowledgment}
D. Liu, R. Wei and L. Han acknowledges financial support by the Natural
Science Foundation of the Jiangsu Province Grant No.\ BK20210198.
S. Dong acknowledge financial support by the National Natural
Science Foundation of China (NNSFC) Grant No.\ 11834002.
Computational resources for most calculations have been provided by
the Michigan State University High Performance Computing Center.


\addcontentsline{toc}{chapter}{References}
\begin{thebibliography}{99}\footnotesize
\itemsep=-3pt plus.2pt minus.2pt   

\bibitem{Triscone07} Rabe K M, Ahn C and Triscone J M 2007
    \href{https://archive-ouverte.unige.ch/unige:12885}
    \emph{$\it{Physics~of~ferroelectrics:~a~modern~perspective}$} (Berlin: Springer)
\bibitem{Eom2004} Choi K J, Biegalski M, Li Y L, Sharan A,
          Schubert J, Uecker R, Reiche P, Chen Y B, Pan, X Q, Gopalan V, Chen L-Q, Schlom D G,
          Eom C B
    \href{https://doi.org/10.1126/science.1103218}{2004
    \emph{Science} \textbf{306} 1005}
\bibitem{Yang2016} Fei R, Kang W and Yang, L.
    \href{https://doi.org/10.1103/PhysRevLett.117.097601}{2016
    \emph{Phys. Rev. Lett.} \textbf{117} 097601}
\bibitem{Xue2017} Tian Z, Guo C, Zhao M, Li R and Xue J
    \href{https://doi.org/10.1021/acsnano.6b08704}{2017
    \emph{ACS Nano} \textbf{11} 2219}
\bibitem{Wang2017} Wang H and Qian X
    \href{https://doi.org/10.1088/2053-1583/4/1/015042}{2017
    \emph{2D Mater.} \textbf{4} 015042}
\bibitem{Duan2020} Guan Z, Hu H, Shen X, Xiang P, Zhong N, Chu J and Duan C
    \href{https://doi.org/10.1002/aelm.201900818}{2020
    \emph{Adv. Elec. Mater.} \textbf{6} 1900818}
\bibitem{Zhu2017} Ding W, Zhu J, Wang Z, Gao Y, Xiao D, Gu Y, Zhang Z and Zhu W
    \href{https://doi.org/10.1038/ncomms14956}{2017
    \emph{Nat. Comm.} \textbf{8} 14956}
\bibitem{Liu2016} Liu F, You L, Seyler K L, Li X, Yu P, Lin J, Wang X, Zhou J, Wang H,
          He H, Pantelides S T, Zhou W, Sharma P, Xu X, Ajayan P M, Wang J and Liu Z
    \href{https://doi.org/10.1038/ncomms12357}{2016
    \emph{Nat. Comm.} \textbf{7} 12357}
\bibitem{Xiao2018} Xiao J, Zhu H, Wang Y, Feng W, Hu Y, Dasgupta A, Han Y,
                   Wang Y, Muller D A, Martin L W, Hu P A and Zhang X
    \href{https://doi.org/10.1103/PhysRevLett.120.227601}{2018
    \emph{Phys. Rev. Lett.} \textbf{120} 227601}
\bibitem{Wu2017} Li L and Wu M
    \href{https://doi.org/10.1021/acsnano.7b02756}{2017
    \emph{ACS Nano} \textbf{11} 6382}
\bibitem{Dong2021} Ding N, Chen J, Gui C, You H, Yao X and Dong S
    \href{https://link.aps.org/doi/10.1103/PhysRevMaterials.5.084405}{2021
    \emph{Phys. Rev. Materials} \textbf{5} 084405}
\bibitem{Guan2021} Song S, Zhang Y, Guan J and Dong S
    \href{https://doi.org/10.1103/PhysRevB.103.L140104}{2021
    \emph{Phys. Rev. B} \textbf{103} L140104}
\bibitem{Ziwen2021} Wang Z, Ding N, Gui C, Wang S, An M and Dong S
    \href{https://doi.org/10.1103/PhysRevMaterials.5.074408}{2021
    \emph{Phys. Rev. Materials} \textbf{5} 074408}
\bibitem{Ding2020} Ding N, Chen J, Dong S and Stroppa A
    \href{https://doi.org/10.1103/PhysRevB.102.165129}{2020
    \emph{Phys. Rev. B} \textbf{102} 165129}
\bibitem{Lin2019} Lin L, Zhang Y, Moreo A, Dagotto E and Dong S
    \href{https://doi.org/10.1103/PhysRevLett.123.067601}{2019
    \emph{Phys. Rev. Lett.} \textbf{123} 067601}
\bibitem{You2019} You Lu, Zhang Y, Zhou S, Chaturvedi A, Morris S A, Liu F,
      Chang L, Ichinose D, Funakubo H, Hu W, Wu T, Liu Z, Dong S and Wang J
    \href{https://doi.org/10.1126/sciadv.aav3780}{2019
    \emph{Sci. Adv.} \textbf{5} eaav3780}
\bibitem{Wu2013} Wu M, Burton J D, Tsymbal E Y, Zeng X C and Jena P
    \href{https://doi.org/10.1103/PhysRevB.87.081406}{2013
    \emph{Phys. Rev. B} \textbf{87} 081406(R)}
\bibitem{Kan2013} Kan E, Wu F, Deng K and Tang W
    \href{https://doi.org/10.1063/1.4829268}{2013
    \emph{Appl. Phys. Lett.} \textbf{103} 193103}
\bibitem{Lin20191} Lin L, Zhang Y, Moreo A, Dagotto E and Dong S
    \href{https://doi.org/10.1103/PhysRevMaterials.3.111401}{2019
    \emph{Phys. Rev. Materials} \textbf{3} 111401(R)}
\bibitem{Hernandez12} Hernandez Y, Pang S, Feng X and Mullen K
    \href{https://doi.org/10.1016/B978-0-444-53349-4.00216-8}{2012
    \emph{Elsevier} 415}
\bibitem{Iijima93} Iijima S and Ichihashi T
    \href{https://doi.org/10.1038/364737d0}{1993
    \emph{Nature} \textbf{363} 603}
\bibitem{Iijima91} Iijima S
    \href{https://doi.org/10.1038/354056a0}{1991
    \emph{Nature} \textbf{354} 56}
\bibitem{Ren2021} Ren Y and Wu M
    \href{https://doi.org/10.1063/5.0035745}{2021
    \emph{J. Chem. Phys.} \textbf{154} 044705}
\bibitem{Yakob2019} Zhang J, Guan J, Dong S and Yakobson B I
    \href{https://doi.org/10.1021/jacs.9b03201}{2019
    \emph{J. Am. Chem. Soc.} \textbf{141} 15040}
\bibitem{Du2021} Zhang L, Tang C, Sanvito S and Du A
    \href{https://doi.org/10.1038/s41524-021-00602-9}{2021
    \emph{Npj Comput. Mater.} \textbf{7} 135}
\bibitem{Zheng2011} Zhao H, Kong X, Li H, Jin Y, Long L, Zeng X, Huang R and Zheng L
    \href{https://doi.org/10.1073/pnas.1010310108}{2011
    \emph{Proc. Natl. Acad. Sci. U.S.A.} \textbf{108} 3481}
\bibitem{Nilges2016} Pfister D, Schafer K, Ott C, Gerke B, Pottgen R,
                     Janka O, Baumgartner M, Efimova A, Hohmann A, Schmidt P,
                     Venkatachalam S, van Wullen L, Schurmann U, Kienle L, Duppel V,
                     Parzinger E, Miller B, Becker J, Holleitner A, Weihrich R and Nilges T
    \href{https://doi.org/10.1002/adma.201603135}{2016
    \emph{Adv. Mater.} \textbf{28} 9783}
\bibitem{Hoff2021} Hoff D A and Rego L G C
    \href{https://doi.org/10.1021/acs.nanolett.1c02636}{2021
    \emph{Nano Lett.} \textbf{21} 8190}
\bibitem{Artacho2008} Artacho E, Anglada E, Dieguez O, Gale J D, Garcia A,
                      Junquera J, Martin R M, Ordejon P, Pruneda J M,
                      Sanchez-Portal D and Soler J M
    \href{https://doi.org/10.1088/0953-8984/20/6/064208}{2008
    \emph{J. Phys.: Condens. Matter} \textbf{20} 064208}
\bibitem{Perdew1996} Perdew J P, Burke K and Ernzerhof M
    \href{https://doi.org/10.1103/PhysRevLett.77.3865}{1996
    \emph{Phys. Rev. Lett.} \textbf{77} 3865}
\bibitem{Troullier1991} Troullier N and Martins J L
    \href{https://doi.org/10.1103/PhysRevB.43.1993}{1991
    \emph{Phys. Rev. B} \textbf{43} 1993}
\bibitem{Pack1976} Monkhorst H J and Pack J D
    \href{https://doi.org/10.1103/PhysRevB.13.5188}{1976
    \emph{Phys. Rev. B} \textbf{13} 5188}
\bibitem{CGmethod} Hestenes M R and Stiefel E
    \href{http://dx.doi.org/10.6028/jres.049.044}{1952
    \emph{J. Res. Natl. Bur. Stand.} \textbf{49} 409}
\bibitem{Resta1993} Resta R, Posternak M and Baldereschi A
    \href{https://doi.org/10.1103/PhysRevLett.70.1010}{1993
    \emph{Phys. Rev. Lett.} \textbf{70} 1010}
\bibitem{Resta2007} Resta R and Vanderbilt D
    \href{https://doi.org/10.1007/978-3-540-34591-6_2}{2007
    \emph{Springer Berlin Heidelberg} 31}
\bibitem{Nilges2016} Pfister D, Schafer K, Ott C, Gerke B, Pottgen R,
                     Janka O, Baumgartner M, Efimova A, Hohmann A,
                     Schmidt P, Venkatachalam S, van Wullen L, Schurmann U,
                     Kienle L, Duppel V, Parzinger E, Miller B, Becker J,
                     Holleitner A, Weihrich R and Nilges T
    \href{https://doi.org/10.1002/adma.201603135}{2016
    \emph{Adv. Mater.} \textbf{28} 9783}
\bibitem{Ren2004} Ren X
    \href{https://doi.org/10.1038/nmat1051}{2004
    \emph{Nat. Mater.} \textbf{3} 91}
\bibitem{Santra2007} Santra B, Michaelides A and Scheffler M
    \href{https://doi.org/10.1063/1.2790009}{2007
    \emph{J. Chem. Phys.} \textbf{127} 184104}
\bibitem{Gillan2016} Gillan M J, Alfe D and Michaelides A
    \href{https://doi.org/10.1063/1.4944633}{2016
    \emph{J. Chem. Phys.} \textbf{144} 130901}
\bibitem{Kyuho07} Lee K, Yu J and Morikawa Y
    \href{https://doi.org/10.1103/PhysRevB.75.045402}{2007
    \emph{Phys. Rev. B} \textbf{75} 045402}
\bibitem{DT256} Liu D, Guan J, Jiang J, and Tomanek D
    \href{https://doi.org/10.1021/acs.nanolett.6b04128}{2016
    \emph{Nano Lett.} \textbf{16} 7865}
\bibitem{Ceperley1980} Ceperley D M and Alder B J
    \href{https://doi.org/10.1103/PhysRevLett.45.566}{2016
    \emph{Phys. Rev. Lett.} \textbf{45} 566}
\end{thebibliography}
\end{document}